\documentclass[aps,pra,final,10pt,twocolumn,superscriptaddress]{revtex4-1}
\usepackage{amsmath,amssymb}
\usepackage{graphicx,color}
\usepackage[squaren]{SIunits}
\usepackage[english]{babel}

\makeatletter
\def\frutiger{cmss10 }
\def\frutigerbold{cmssbx10 }
=\frutigerbold at 14pt
=\frutiger at 8pt
=\frutigerbold at 12pt
=\frutigerbold at10pt
=\frutigerbold at 8pt
=\frutiger at 8pt
=\frutigerbold at 31pt
\def\@caption@tabnum@sep{\figtextfont{{ }{\bf\textbar}{ }}}%
\def\fnum@table{{\bf\tablename~\thetable}}
\renewenvironment{table}{\@float{table}\def\textbf##1{{\fignumfont ##1}}\def\bf{\fignumfont}}{\end@float}
\def\@caption@fignum@sep{\figtextfont{{ }{\bf\textbar}{ }}}%
\def\fnum@figure{{\bf\figurename~\thefigure}}
\renewenvironment{figure}{\@float{figure}\def\textbf##1{{\fignumfont ##1}}\def\bf{\fignumfont}}{\end@float}
\def\@startsection#1#2#3#4#5#6{%
\if@noskipsec\leavevmode\fi
\par\@tempskipa #4\relax
\@afterindenttrue
\ifdim\@tempskipa <\z@
\@tempskipa -\@tempskipa \@afterindentfalse
\fi\if@nobreak\everypar{}%
\else\addpenalty\@secpenalty\addvspace\@tempskipa\fi
\@ifstar{\@ssect{#3}{#4}{#5}{#6}}{\@dblarg{\@sect{#1}{#2}{#3}{#4}{#5}{#6}}}}
\def\@sect#1#2#3#4#5#6[#7]#8{%
\ifnum #2>0
\let\@svsec\@empty
\else\refstepcounter{#1}\protected@edef\@svsec{\@seccntformat{#1}\relax}\fi
\@tempskipa #5\relax
\ifdim\@tempskipa>\z@
\begingroup#6{\@hangfrom{\hskip #3\relax\@svsec}%
\interlinepenalty \@M #8\@@par}\endgroup
\csname #1mark\endcsname{#7}%
\addcontentsline{toc}{#1}{%
\ifnum #2>\c@secnumdepth\else
\protect\numberline{\csname the#1\endcsname}\fi #7}%
\else\def\@svsechd{#6{\hskip #3\relax
\@svsec #8\ifnum#2=2.\fi}%
\csname #1mark\endcsname{#7}%
\addcontentsline{toc}{#1}{%
\ifnum #2>\c@secnumdepth \else
\protect\numberline{\csname the#1\endcsname}\fi #7}}%
\fi\@xsect{#5}}
\renewcommand\section{\@startsection {section}{1}{\z@}%
{-10pt \@plus -1ex \@minus -.2ex}{.5ex }{\normalfont\Large\bfseries\sectionfont}}
\renewcommand\subsection{\@startsection{subsection}{2}{\z@}%
{10pt\@plus 1ex \@minus .2ex}{-0.5ex \@plus .2ex}{\normalfont\large\bfseries\subsectionfont}}
\def\frontmatter@title@format{\titlefont\centering}%
\def\frontmatter@title@below{\addvspace{-5pt}}%
\def\dropcap#1{\setbox1=\hbox{\dropcapfont\uppercase{#1}\hskip1pt}
\hangindent=\wd1 \hangafter-2 \noindent\llap{\vbox to0pt{\vskip-7pt\copy1\vss}}}
\renewenvironment{thebibliography}[1]{%
\bib@heading%
  \ifx\bibpreamble\relax\else\ifx\bibpreamble\@empty\else
    \noindent\bibpreamble\par\nobreak
  \fi\fi
  \list{\@biblabel{\@arabic\c@enumiv}}%
  {\settowidth\labelwidth{\@biblabel{#1}}%
    \leftmargin\labelwidth
    \advance\leftmargin\labelsep
    \@openbib@code
    \usecounter{enumiv}%
    \let\p@enumiv\@empty
    \renewcommand*\theenumiv{\@arabic\c@enumiv}
}%
  \sloppy\clubpenalty4000\widowpenalty4000%
  \sfcode`\.=\@m}
{\def\@noitemerr
  {\@latex@warning{Empty `thebibliography' environment}}%
  \endlist}
\newcommand*\bib@heading{%
  \section{\refname}
  \fontsize{8}{10}\selectfont
}
\newcommand*\@openbib@code{%
      \advance\leftmargin\bibindent
      \itemindent -\bibindent
      \listparindent \itemindent
      \parsep \z@
}%
\newdimen\bibindent
\makeatother

\newcommand{\DT}{\mathbf{\underline{D}}_{\mathrm{T}}}%
\newcommand{\AU}[1]{{\fignumfont #1}}%
\newcommand{\dif}{\mathrm{d}}%

\newcommand{\breite}{\linewidth}%

\begin{document}

\noindent Nature Communications \textbf{5}, 4829 (2014) \\ 
DOI: 10.1038/ncomms5829 \\
\url{http://www.nature.com/ncomms/2014/140919/ncomms5829/full/ncomms5829.html}
\vspace{0.6cm}

\title{Gravitaxis of asymmetric self-propelled colloidal particles}  

\author{Borge ten Hagen}
\affiliation{Institut f{\"u}r Theoretische Physik II: Weiche Materie,
Heinrich-Heine-Universit{\"a}t D{\"u}sseldorf, D-40225 D{\"u}sseldorf, Germany}

\author{Felix K{\"u}mmel}
\affiliation{2.\ Physikalisches Institut, Universit{\"a}t Stuttgart, D-70569 Stuttgart, Germany} 

\author{Raphael Wittkowski}
\affiliation{SUPA, School of Physics and Astronomy, University of Edinburgh, Edinburgh, EH9 3JZ, United Kingdom}

\author{Daisuke Takagi}
\affiliation{Department of Mathematics, University of Hawaii at Manoa, Honolulu, Hawaii
96822, USA}

\author{Hartmut L{\"o}wen}
\affiliation{Institut f{\"u}r Theoretische Physik II: Weiche Materie,
Heinrich-Heine-Universit{\"a}t D{\"u}sseldorf, D-40225 D{\"u}sseldorf, Germany}

\author{Clemens Bechinger}
\email{Electronic address: c.bechinger@physik.uni-stuttgart.de} 
\affiliation{2.\ Physikalisches Institut, Universit{\"a}t Stuttgart, D-70569 Stuttgart, Germany} 
\affiliation{Max-Planck-Institut f{\"u}r Intelligente Systeme, D-70569 Stuttgart, Germany}

\date{\today}

\begin{abstract}
Many motile microorganisms adjust their swimming motion relative to the gravitational field and thus counteract sedimentation to the ground. This gravitactic behavior is often the result of an inhomogeneous mass distribution which aligns the microorganism similar to a buoy. However, it has been suggested that gravitaxis can also result from a geometric fore-rear asymmetry, typical for many self-propelling organisms. Despite several attempts, no conclusive evidence for such an asymmetry-induced gravitactic motion exists. Here, we study the motion of asymmetric self-propelled colloidal particles which have a homogeneous mass density and a well-defined shape. In experiments and by theoretical modeling we demonstrate that a shape anisotropy alone is sufficient to induce gravitactic motion with either preferential upward or downward swimming. In addition, also trochoid-like trajectories transversal to the direction of gravity are observed.
\end{abstract}

\maketitle


\dropcap{G}ravitaxis (known historically as geotaxis) describes the response of motile microorganisms to an external gravitational field and has been studied for several decades. In particular, negative gravitaxis, i.e., a swimming motion opposed to a gravitational force $\mathbf{F}_{\mathrm{G}}$, is frequently observed for flagellates and ciliates such as \emph{Chlamydomonas} \cite{roberts2006mechanisms} or \emph{Paramecium} \cite{roberts2010mechanics}. Since this ability enables microorganisms to counteract sedimentation, gravitaxis extends the range of their habitat and allows them to optimize their position in environments with spatial gradients \cite{roberts2010mechanics}. In case of photosynthetic flagellates such as \textit{Euglena gracilis}, gravitaxis (in addition to phototaxis) contributes to their vertical motion in water which enables them to adjust the amount of exposure to solar radiation \cite{ntefidou2003photoactivated}. In order to achieve a gravitactic motion, in general a stable orientation of the microorganism relative to the gravitational field is required. While in some organisms such alignment is likely to be dominated by
an active physiological mechanism \cite{hader1997graviperception,hader2009molecular}, its origin in other systems is still controversially discussed \cite{richter2007gravitaxis,hemmersbach1999graviorientation,hader1999gravitaxis,nagel2000physical,machemer1996gravitaxis,roberts2002gravitaxis,roberts2006mechanisms, roberts2010mechanics,mogami2001theoretical,streb2002sensory}. It has been suggested that gravitaxis can also result from purely passive effects like, e.g., an inhomogeneous mass density within the organism (bottom-heaviness), which would lead to an alignment similar to that of a buoy in the ocean \cite{durham2009disruption, kessler1985hydrodynamic}. 
(We use the term ``gravitaxis'', which is not uniformly defined in the literature, also in the context of passive alignment effects.) 
Under such conditions, the organism's alignment should be the same regardless whether it sediments downward or upward in a hypo- or hyper-density medium, respectively, as indeed observed for \textit{pluteus larvae} \cite{mogami2001theoretical}. However, corresponding experiments with immobilized \textit{Paramecium} and \textit{gastrula larvae} showed an alignment reversal for opposite sedimentation directions \cite{mogami2001theoretical}. This suggests another mechanical alignment mechanism which is caused by the organism's fore-rear asymmetry \cite{roberts2002gravitaxis,roberts2006mechanisms, roberts2010mechanics,mogami2001theoretical}.

To explore the role of shape asymmetry on the gravitactic behavior and to avoid the presence of additional physiological mechanisms, in our experiments we study the motion of colloidal micron-sized swimmers with asymmetric shapes in a gravitational field. 
The self-propulsion mechanism is based on diffusiophoresis, where a local chemical gradient is induced in the solvent around the microswimmer \cite{anderson1989colloid}. It has been shown that the corresponding type of motion is similar to that of biological microorganisms \cite{howse2007self,hong2007chemotaxis,gibbs2009autonomously,jiang2010active,golestanian2005propulsion,theurkauff2012dynamic,PalacciPRL2010}. 
Based on our experimental and theoretical results, we demonstrate that a shape anisotropy alone is sufficient to induce gravitactic motion with either upward or downward swimming. In addition to straight trajectories also more complex swimming patterns are observed, where the swimmers perform a trochoid-like (i.e., a generalized cycloid-like \cite{Yates}) motion transversal to the direction of gravity.

\section{Results}
\subsection{Experiments}
For our experiments we use asymmetric L-shaped microswimmers with arm lengths of 9 and \unit{6}{\micro\metre}, respectively, and \unit{3}{\micro\metre} thickness that are obtained by soft lithography \cite{badaire2007shape,kummel2013circular} (see the Methods section for details). To induce a self-diffusiophoretic motion, the particles are covered with a thin Au coating on the front side of the short arm, which leads to local heating upon laser illumination with intensity $I$ [see Fig.\ \ref{Fig.1}(d)]. When such particles are suspended in a binary mixture of water and 2,6-lutidine at critical composition, this heating causes a local demixing of the solvent which results in an intensity-dependent phoretic propulsion in the direction normal to the plane of the metal cap \cite{kummel2013circular,volpe2011microswimmers,buttinoni2012active}. 
To restrict the particle's motion to two spatial dimensions, we use a sample cell with a height of \unit{7}{\micro\metre}. Further details are provided in the Methods section. Variation of the gravitational force is achieved by mounting the sample cell on a microscope which can be inclined by an angle $\alpha$ relative to the horizontal plane [see Fig.\ \ref{Fig.1}(b)].

\subsection{Passive sedimentation}
\begin{figure}
\centering
\includegraphics[width=\breite]{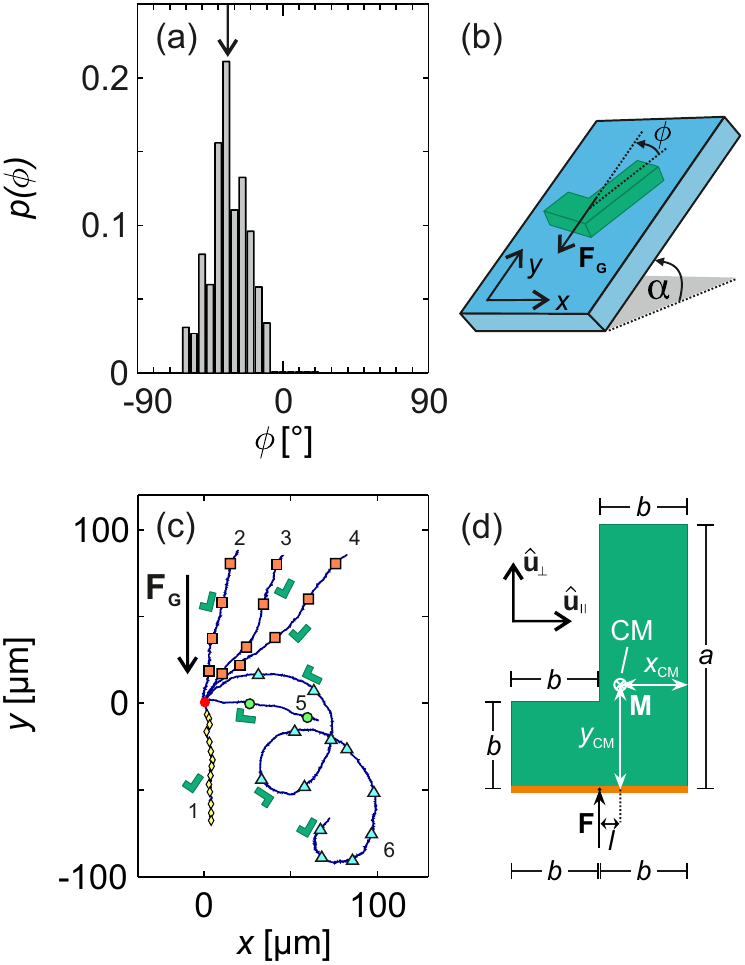}
\caption{\label{Fig.1}\AU{Characterization of the experimental setup.} (\textbf{a}) Measured probability distribution $p(\phi)$ of the orientation $\phi$ of a passive L-shaped particle during sedimentation for an inclination angle $\alpha=\unit{10.67}{\degree}$ [see sketch in (\textbf{b})]. (\textbf{c}) Experimental trajectories for the same inclination angle and increasing illumination intensity: (1) $I=0$, (2-5) $\unit{0.6}{\micro\watt\micro\rpsquare\metre} < I < \unit{4.8}{\micro\watt\micro\rpsquare\metre}$, (6) $I > \unit{4.8}{\micro\watt\micro\rpsquare\metre}$.
All trajectories start at the origin of the graph (red bullet). The particle positions after \unit{1}{\minute} each are marked by yellow diamonds (passive straight downward trajectory), orange squares (straight upward trajectories), green bullets (active tilted straight downward trajectory), and blue triangles (trochoid-like trajectory). (\textbf{d}) Geometrical sketch of an ideal L-shaped particle (the Au coating is indicated by the yellow line) with dimensions $a=\unit{9}{\micro\metre}$ and $b=\unit{3}{\micro\metre}$ and coordinates $x_{\mathrm{CM}}=\unit{-2.25}{\micro\metre}$ and $y_{\mathrm{CM}}=\unit{3.75}{\micro\metre}$ of the center of mass CM (with the origin of coordinates in the bottom right corner of the particle and when the Au coating is neglected). The propulsion mechanism characterized by the effective force $\mathbf{F}$ and the effective torque $\mathbf{M}$ induces a rotation of the particle that depends on the length of the effective lever arm $l$. $\mathbf{\hat{u}}_{\parallel}(\phi)$ and $\mathbf{\hat{u}}_{\perp}(\phi)$ are particle-fixed unit vectors that denote the orientation of the particle (see the Methods section for details).}
\end{figure}

Figure \ref{Fig.1}(a) shows the measured orientational probability distribution $p(\phi)$ of a sedimenting passive L-shaped particle ($I=0$) in a thin sample cell that was tilted by $\alpha=\unit{10.67}{\degree}$ relative to the horizontal plane. The data show a clear maximum at the orientation angle $\phi=\unit{-34}{\degree}$, i.e., the swimmer aligns slightly turned relative to the direction of gravity as schematically illustrated in Fig.\ \ref{Fig.1}(b). A typical trajectory for such a sedimenting particle is plotted in Fig.\ \ref{Fig.1}(c),1.
To estimate the effect of the Au layer on the particle orientation, we also performed sedimentation experiments for non-coated particles and did not find measurable deviations. This suggests that the alignment cannot be attributed to an inhomogeneous mass distribution.
The observed alignment with the \emph{shorter} arm at the bottom [see Fig.\ \ref{Fig.1}(b)] is characteristic for sedimenting objects with homogeneous mass distribution and a fore-rear asymmetry \cite{roberts2002gravitaxis}. This can easily be understood by considering an asymmetric dumbbell formed by two spheres with identical mass density but different radii $R_{1}$ and $R_{2}>R_{1}$. The sedimentation speed of a single sphere with radius $R$ due to gravitational and viscous forces is $v \propto R^{2}$. Therefore, if hydrodynamic interactions between the spheres are ignored, the dumbbell experiences a viscous torque resulting in an alignment where the bigger sphere is below the smaller one \cite{roberts2002gravitaxis}.

In Fig.\ \ref{Fig.1}(c) we show the particle's center-of-mass motion in the $x$-$y$ plane as a consequence of the self-propulsion  acting normally to the Au coating [see Fig.\ \ref{Fig.1}(d)]. When the self-propulsion is sufficiently high to overcome sedimentation, the particle performs a rather rectilinear motion in upward direction, i.e., against gravity (negative gravitaxis \cite{roberts2002gravitaxis}, see Fig.\ \ref{Fig.1}(c),2-4). With increasing self-propulsion, i.e., light intensity, the angle between the trajectory and the $y$ axis increases until it exceeds $\unit{90}{\degree}$. For even higher intensities, interestingly, the particle performs an effective downward motion again (see Fig.\ \ref{Fig.1}(c),5). It should be mentioned that in case of the above straight trajectories, the particle's orientation $\phi$ remains stable (apart from slight fluctuations) and shows a monotonic dependence on the illumination intensity (as will be discussed in more detail further below). 
Finally, for strong self-propulsion corresponding to high light intensity, the microswimmer performs a trochoid-like motion (see Fig.\ \ref{Fig.1}(c),6).

\begin{figure}
\centering
\includegraphics[width=\breite]{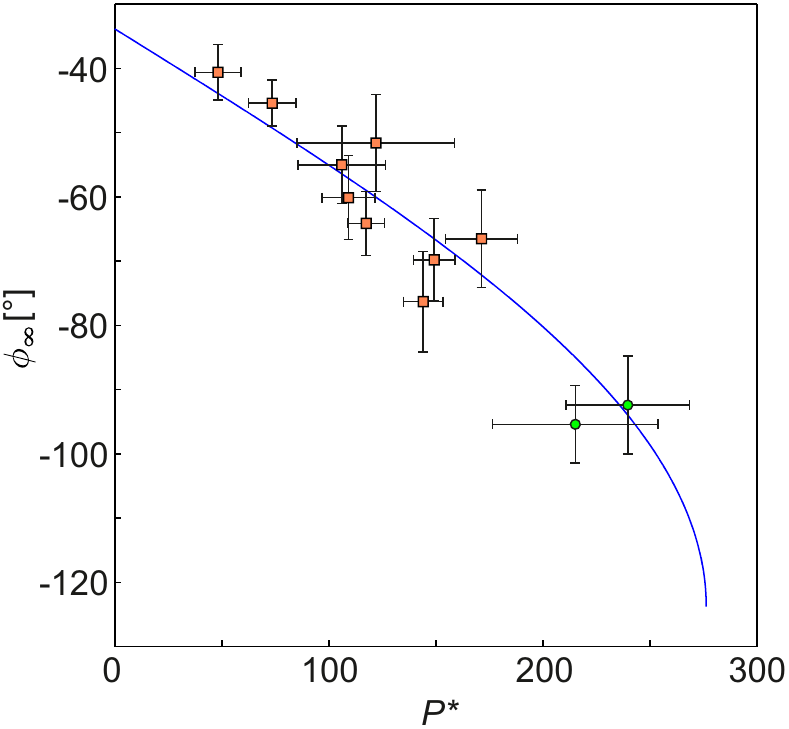}
\caption{\label{Fig.2}\AU{Measured particle orientations.} Long-time orientation $\phi_{\infty}$ of an L-shaped particle in the regime of straight motion as a function of the strength $P^{*}$ of the self-propulsion for $\alpha=\unit{10.67}{\degree}$. 
The symbols with error bars representing the standard deviation show the experimental data with the self-propulsion strength $P^{*}$ determined using eq.\ \eqref{eq:F} and orange squares and green bullets corresponding to upward and downward motion, respectively [see Fig.\ \ref{Fig.1}(c)]. The solid curve is the theoretical prediction based on eq.\ \eqref{eq:phipro} with $D^{\parallel}_{\mathrm{C}}$ as fit parameter (see the Methods section for details).}
\end{figure}

To obtain a theoretical understanding of gravitaxis of asymmetric microswimmers, we first study the sedimentation of \emph{passive} particles in a viscous solvent \cite{Brenner72} based on the Langevin equations
\begin{equation}
\begin{split}
\dot{\mathbf{r}}&=\beta \DT \mathbf{F}_{\mathrm{G}} + \boldsymbol{\zeta}_{\mathbf{r}} \;,\\
\dot{\phi}&= \beta \mathbf{D}_{\mathrm{C}}\!\cdot\!\mathbf{F}_{\mathrm{G}} + \zeta_{\phi}  
\end{split}
\label{eq:passiveLangevin}%
\end{equation}
for the time-dependent center-of-mass position $\mathbf{r}(t)=(x(t),y(t))$ and orientation $\phi(t)$ of a particle. 
Here, the gravitational force $\mathbf{F}_{\mathrm{G}}$, the translational short-time diffusion tensor $\DT $, the translational-rotational coupling vector $\mathbf{D}_{\mathrm{C}}$, the inverse effective thermal energy $\beta=1/(k_{\mathrm{B}}T)$, and the Brownian noise terms $\boldsymbol{\zeta}_{\mathbf{r}}$ and $\zeta_{\phi}$ are involved (see the Methods section for details). 
 
Neglecting the stochastic contributions in eqs.\ \eqref{eq:passiveLangevin}, one obtains the stable long-time particle orientation angle
\begin{equation}
\phi_{\infty} \equiv \phi(t \rightarrow \infty) = -\arctan\!\bigg(\frac{D^{\perp}_{\mathrm{C}}}{D^{\parallel}_{\mathrm{C}}}\bigg)\,,
\label{eq:phigra}%
\end{equation}
which only depends on the two coupling coefficients $D^{\parallel}_{\mathrm{C}}$ and
$D^{\perp}_{\mathrm{C}}$ determined by the particle's geometry.

\subsection{Swimming patterns under gravity}
Extending eqs.\ \eqref{eq:passiveLangevin} to also account for the \emph{active} motion of an asymmetric micro\-swimmer yields (see the Methods section for a hydrodynamic derivation)
\begin{align}%
\dot{\mathbf{r}}&= (P^{*}/b) \big(\DT \mathbf{\hat{u}}_{\perp}+l\mathbf{D}_{\mathrm{C}} \big)
+ \beta \DT \mathbf{F}_{\mathrm{G}}   + \boldsymbol{\zeta}_{\mathbf{r}} \;,\label{eq:Langevinr}\\
\dot{\phi}&= (P^{*}/b) \big(l D_{\mathrm{R}} +\mathbf{D}_{\mathrm{C}}\!\cdot\!\mathbf{\hat{u}}_{\perp}\big) + \beta \mathbf{D}_{\mathrm{C}}\!\cdot\!\mathbf{F}_{\mathrm{G}} + \zeta_{\phi} \,,
\label{eq:Langevinphi}%
\end{align}%
where the dimensionless number $P^{*}$ is the strength of the
self-propulsion, $b$ is a characteristic length of the L-shaped particle [see Fig.\ \ref{Fig.1}(d)], $l$ denotes an effective lever arm [see Fig.\ \ref{Fig.1}(d)], and $D_{\mathrm{R}}$ is the rotational diffusion coefficient of the particle.
As shown in the Methods section, one can view $F=\lvert\mathbf{F}\rvert=P^{*}/(b \beta)$ and  $M=\lvert\mathbf{M}\rvert=Fl$ as an 
effective force (in $\mathbf{\hat{u}}_{\perp}$-direction) and an effective
torque (perpendicular to the L-shaped particle), respectively, describing the self-propulsion of the particle [see Fig.\ \ref{Fig.1}(d)]. This concept is in line with other theoretical work that has been presented recently \cite{Friedrich2008,JekelyEtAl2008,Radtke2012,Volpe2013,Crespi2013,Marine2013}.
The above equations of motion are
fully compatible with the fact that, apart from gravity, a self-propelled swimmer
is force-free and torque-free (see the Methods section).

The self-propulsion strength $P^{*}$ is obtained from the experiments
by measuring the particle velocity (see the Methods section for details). The rotational motion of the microswimmer depends on the detailed asymmetry of the particle shape and is characterized by the
effective lever arm $l$ relative to the center-of-mass position as the reference point [see Fig.\ \ref{Fig.1}(d)].
Note that the choice of the reference point also changes the translational and coupling elements of the diffusion tensor. If the reference point does not coincide with the center of mass of the particle, in the presence of a gravitational force an additional torque has to be considered in eq.\ \eqref{eq:Langevinphi}.
In line with our experiments [see Fig.\ \ref{Fig.1}(c)], the noise-free asymptotic solutions of  eqs.\ \eqref{eq:Langevinr} and \eqref{eq:Langevinphi} are either straight (upward or downward) trajectories or periodic swimming paths. Up to a threshold value of $P^{*}$, the effective torque originating from the self-propulsion [first term on the right-hand-side of eq.\ \eqref{eq:Langevinphi}] can be compensated by the gravitational torque [second term $\beta \mathbf{D}_{\mathrm{C}}\!\cdot\!\mathbf{F}_{\mathrm{G}}$ on the right-hand-side of eq.\ \eqref{eq:Langevinphi}] so that there is no net rotation and the trajectories are straight. In this case, after a transient regime the particle orientation converges to
\begin{equation}
\phi_{\infty} = - \arctan\!\bigg(\frac{D^{\perp}_{\mathrm{C}}}{D^{\parallel}_{\mathrm{C}}}\bigg) 
+ \arcsin\!\Bigg(\!\!-\frac{P^{*}}{\beta b F_{\mathrm{G}}} \frac{D^{\perp}_{\mathrm{C}}+l D_{\mathrm{R}}}{\sqrt{D^{\parallel 2}_{\mathrm{C}} 
+ D^{\perp 2}_{\mathrm{C}}}}\Bigg)
\label{eq:phipro}%
\end{equation}
with the gravitational force $F_{\mathrm{G}}=\lvert\mathbf{F}_{\mathrm{G}}\rvert$. Obviously, $\phi_{\infty}$ is a superposition of the passive case [first term on the right-hand-side, cf.\ eq.\ \eqref{eq:phigra}] with a correction due to self-propulsion (second term on the right-hand-side). The theoretical prediction given by eq.\ \eqref{eq:phipro} is visualized in Fig.\ \ref{Fig.2} by the solid line, which is fitted to the experimental data (symbols) by using $D^{\parallel}_{\mathrm{C}}$ as fit parameter.

The restoring torque caused by gravity depends on the orientation of the particle and becomes maximal at the critical angle $\phi_{\mathrm{crit}} = \arctan (D^{\parallel}_{\mathrm{C}}/D^{\perp}_{\mathrm{C}})-\pi$. According to eq.\ \eqref{eq:phipro}, this corresponds to a critical self-propulsion strength
\begin{equation}
P^{*}_\mathrm{crit} = \beta b mg \sin \alpha \frac{\sqrt{D^{\parallel 2}_{\mathrm{C}} 
+ D^{\perp 2}_{\mathrm{C}}}}{D^{\perp}_{\mathrm{C}}+ l D_{\mathrm{R}}} 
\label{eq:Fcrit}%
\end{equation}
with the particle's buoyant mass $m$ and the gravity acceleration of earth $g=\unit{9.81}{\metre \rpsquare\second}$.
When $P^{*}$ exceeds this critical value for a given inclination angle of the setup, the effective torque originating from the non-central drive can no longer be compensated by the restoring torque due to gravity [see eq.\ \eqref{eq:Langevinphi}] so that $\dot{\phi}\neq 0$ and a periodic  motion occurs.

To apply our theory to the experiments, we use the values of the various parameters as shown in Table \ref{tab}. All data are obtained from our measurements as described in the Methods section. 

\begin{table}
\caption{\label{tab}Experimentally determined values of the parameters used for the comparison with our theory.}
\begin{ruledtabular}
\begin{tabular}{lcc}
translational diffusion coefficients &  $D_{\parallel}$ & $\unit{7.2 \times 10^{-3}}{\micro \squaren \metre\reciprocal\second}$  \\
& $D_{\perp}$  & $\unit{8.1 \times 10^{-3}}{\micro \squaren \metre\reciprocal\second}$  \\
& $D_{\parallel}^{\perp}$ & $\unit{0}{\micro \squaren \metre\reciprocal\second}$ \\
\hline 
coupling coefficients & $D^{\parallel}_{\mathrm{C}}$ & $\unit{5.7 \times 10^{-4}}{\micro\metre\reciprocal\second}$ \\
& $D^{\perp}_{\mathrm{C}}$ & $\unit{3.8 \times 10^{-4}}{\micro\metre\reciprocal\second}$ \\
\hline
rotational diffusion coefficient & $D_{\mathrm{R}}$ & $\unit{6.2 \times 10^{-4}}{\reciprocal\second}$  \\
\hline
effective lever arm & $l$ & $\unit{-0.75}{\micro\metre}$  \\
\hline
buoyant mass & $m$ & $\unit{2.5 \times 10^{-14}}{\kilogram}$  \\
\end{tabular}
\end{ruledtabular}
\end{table}

\begin{figure}
\centering
\includegraphics[width=\breite]{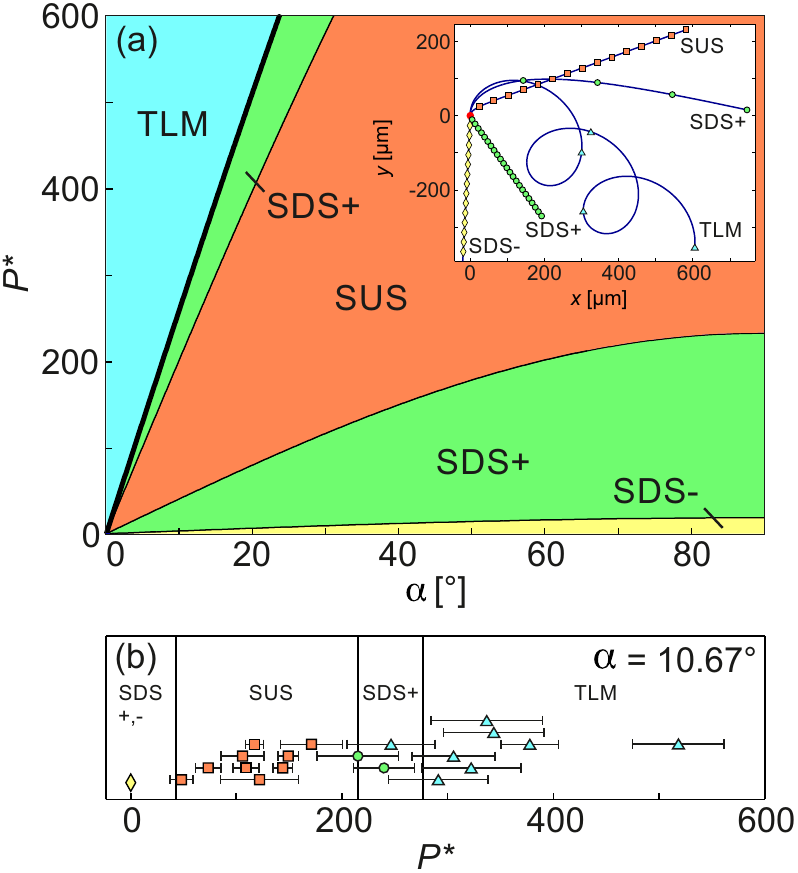}
\caption{\label{Fig.3}\AU{State diagram for moderate effective lever arm.} (\textbf{a}) State diagram of the motion of an active L-shaped particle with an effective lever arm $l=\unit{-0.75}{\micro\metre}$ [see sketch in Fig.\ \ref{Fig.1}(d)] under gravity. The types of motion are straight downward swimming (SDS), straight upward swimming (SUS), and trochoid-like motion (TLM).  Straight downward swimming is usually accompanied by a drift in negative (SDS-) or positive (SDS+) $x$ direction.  
The transition from straight to circling motion is marked by a thick black line and determined analytically by eq.\ \eqref{eq:Fcrit}. 
Theoretical noise-free example trajectories for the various states are shown in the inset. All trajectories start at the origin and the symbols (diamonds: \mbox{SDS-,} circles: SDS+, squares: SUS, triangles: TLM) indicate particle positions after \unit{5}{\minute} each. (\textbf{b}) Experimentally observed types of motion for $\alpha=\unit{10.67}{\degree}$. The different symbols correspond to the various states as defined in (\textbf{a}) and are shifted in vertical direction for clarity. The error bars represent the standard deviation.}
\end{figure}

\subsection{Dynamical state diagram}    
Depending on the self-pro\-pul\-sion strength $P^{*}$ and the substrate inclination angle $\alpha$, different types of motion occur [see Fig.\ \ref{Fig.3}(a)]. (For clarity we neglected the noise in Fig.\ \ref{Fig.3}, but we checked that noise changes the trajectories only marginally.)
For very small values of $P^{*}$ the particle performs straight downward swimming (SDS). The theoretical calculations reveal two regimes where on top of the downward motion either a small drift to the left (SDS-) or to the right (SDS+) is superimposed. 
For zero self-propulsion this additional lateral drift originates from the difference $D_{\parallel}-D_{\perp}$ between the translational diffusion coefficients, which is characteristic for non-spherical particles.
Further increasing $P^{*}$ results in negative gravitaxis, i.e., straight upward swimming (SUS). Here, the vertical component of the self-propulsion counteracting gravity exceeds the strength of the gravitational force. 
For even higher $P^{*}$, the velocity-dependent torque exerted on the particle further increases and leads to trajectories that are tilted more and more until a re-entrance to a straight downward motion with a drift to the right (SDS+) is observed. For the highest values of $P^{*}$ the particle performs a trochoid-like motion (TLM). 
The critical self-propulsion $P^{*}_{\mathrm{crit}}$ for the transition from straight to trochoid-like motion as obtained analytically from eq.\ \eqref{eq:Fcrit} is indicated by a thick black line in Fig.\ \ref{Fig.3}(a).

Indeed, the experimentally observed types of motion taken from Fig.\ \ref{Fig.1}(c) correspond to those in the theoretical state diagram. This is shown in Fig.\ \ref{Fig.3}(b) where we plotted the experimental data for $\alpha=\unit{10.67}{\degree}$ as a function of the self-propulsion strength $P^{*}$. Apart from small deviations due to thermal fluctuations, quantitative agreement between experiment and theory is obtained.

\begin{figure}
\centering
\includegraphics[width=\breite]{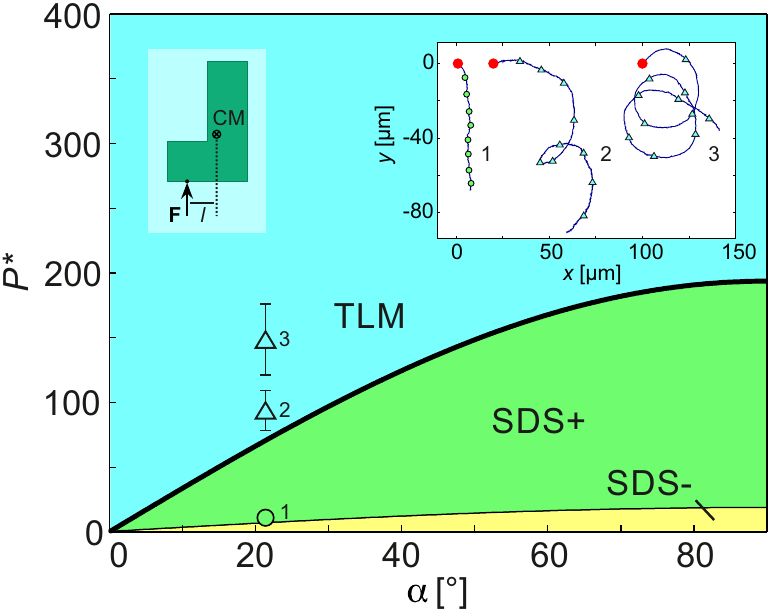}
\caption{\label{Fig.4}\AU{State diagram for large effective lever arm.} The same as in Fig.\ \ref{Fig.3}(a) but now for an active L-shaped particle with effective lever arm $l=\unit{-1.65}{\micro\metre}$ (see left inset). Notice that the regime of upward swimming is missing here. Only straight downward swimming (SDS) [with a drift in negative (SDS-) or positive (SDS+) $x$ direction] and trochoid-like motion (TLM) are observed. The right inset shows three experimental trajectories whose corresponding positions in the state diagram are indicated together with error bars representing the standard deviation.}
\end{figure}

According to eqs.\ \eqref{eq:Langevinr} and \eqref{eq:Langevinphi}, the state diagram should strongly depend on the effective lever arm $l$. To test this prediction experimentally, we tilted the silicon wafer with the L-particles about $\unit{25}{\degree}$ relative to the Au source during the evaporation process. As a result, the Au coating slightly extends over the front face of the L-particles to one of the lateral planes which results in a shift of the effective propulsion force and a change in the lever arm. The value of $l$ was experimentally determined to $l=\unit{-1.65}{\micro\metre}$ from the mean radius of the circular particle motion which is observed for $\alpha=\unit{0}{\degree}$ \cite{kummel2013circular}. The gravitactic behavior of such particles is shown for different self-propulsion strengths in the inset of Fig.\ \ref{Fig.4}. Interestingly, under such conditions we did not find evidence for negative gravitaxis, i.e., straight upward motion. This is in good agreement with the corresponding theoretical state diagram shown in Fig.\ \ref{Fig.4} and suggests that the occurrence of negative gravitaxis does not only depend on the strength of self-propulsion but also on the position where the effective force accounting for the self-propulsion mechanism acts on the body of the swimmer.

\section{Discussion}
Our model allows to derive general criteria for the existence of negative gravitaxis, which are applicable to arbitrary particle shapes. For a continuous upward motion, first, the self-propulsion $P^{*}$ trivially has to be strong enough to overcome gravity. Secondly, the  net rotation of the particle must vanish, which is the case for $P^{*}\leqslant P^{*}_{\mathrm{crit}}$ [see eq.\ \eqref{eq:Fcrit}]. Otherwise, e.g., trochoid-like trajectories are observed.
As shown in the state diagrams in Figs.\ \ref{Fig.3} and \ref{Fig.4}, the motional behavior depends sensitively on several parameters such as the strength of the self-propulsion and the length of the effective lever arm. This may also provide an explanation why gravitaxis is only observed for particular microorganisms as, e.g., \emph{Chlamydomonas reinhardtii} and why the gravitactic behavior is subjected to large variations within a single population \cite{hemmersbach1999graviorientation}.

In our experiments, the particles had an almost homogeneous mass distribution in order to reveal pure shape-induced gravitaxis.
For real biological systems, gravitaxis is a combination of the shape-induced mechanism studied here and bottom-heaviness resulting from an inhomogeneous mass distribution \cite{campbell2013gravitaxis}. 
In particular for eukaryotic cells, small mass inhomogeneities due to the nucleus or organella are rather likely.
Bottom-heaviness can straight-forwardly be included in our modeling by an additional torque contribution \cite{wolff2013sedimentation} and would support negative gravitaxis.
Even for microorganisms with axial symmetry but fore-rear asymmetry \cite{roberts2002gravitaxis} like \emph{Paramecium} \cite{roberts2010mechanics} both the associated shape-dependent hydrodynamic friction and  bottom-heaviness contribute to gravitaxis so that the particle shape matters also in this special case.

In conclusion, our theory and experiments demonstrate that the presence of a gravitational field leads to straight downward and upward motion (gravitaxis) and also to trochoid-like trajectories of asymmetric self-propelled colloidal particles. Based on a set of Langevin equations, one can predict from the shape of a microswimmer, its mass distribution, and the geometry of the self-propulsion mechanism whether negative gravitaxis can occur. Although our study does not rule out additional physiological mechanisms for gravitaxis \cite{hader1997graviperception,hader2009molecular}, our results suggest that a swimming motion of biological microswimmers opposed to gravity can be entirely caused by a fore-rear asymmetry, i.e., passive effects \cite{roberts2002gravitaxis}. Such passive alignment mechanisms may also be useful in situations where directed motion of autonomous self-propelled objects is required as, e.g., in applications where they serve as microshuttles for directed cargo delivery.
Furthermore, their different response to a gravitational field could be utilized to sort microswimmers with respect to their shape and activity \cite{Volpe2013}.

\section{Methods}
\subsection{Experimental details}
Asymmetric L-shaped microswimmers were fabricated from photoresist SU-8 by photolithography. A \unit{3}{\micro\metre} thick layer of SU-8 is spin coated onto a silicon wafer, soft baked for \unit{80}{\second} at \unit{95}{\degreecelsius}, and then exposed to ultraviolet light through a photomask. After a postexposure bake at \unit{95}{\degreecelsius} for \unit{140}{\second} the entire wafer with the attached particles is covered by a several nm thick Au coating. During this process, the substrate is aligned by a specific angle relative to the evaporation source. Depending on the chosen angle, the Au coating can selectively be applied to specific regions of the front sides of the short arms of the particles. Afterwards the particles are released from the substrate by an ultrasonic treatment and suspended in a binary mixture of water and 2,6-lutidine at critical composition (28.6 mass percent of lutidine) that is kept several degrees below its lower critical point $T_{\mathrm{C}} = \unit{34.1}{\degreecelsius}$. Time-dependent particle positions and orientations were acquired by video microscopy at a frame rate of 7.5 fps and stored for further analysis.

\subsection{Langevin equations for a sedimenting passive particle}
The Langevin equations for the center-of-mass position $\mathbf{r}(t)$ and orientation $\phi(t)$ of an arbitrarily shaped passive particle that sediments under the gravitational force $\mathbf{F}_{\mathrm{G}}=(0,-mg \sin \alpha)$ are given by eqs.\ \eqref{eq:passiveLangevin}. The key quantities are the translational short-time diffusion tensor
$\DT =\,D_{\parallel}\mathbf{\hat{u}}_{\parallel}\otimes\mathbf{\hat{u}}_{\parallel}
+D^{\perp}_{\parallel}(\mathbf{\hat{u}}_{\parallel}\otimes\mathbf{\hat{u}}_{\perp}+\mathbf{\hat{u}}_{\perp}\!\otimes\mathbf{\hat{u}}_{\parallel})
+D_{\perp}\mathbf{\hat{u}}_{\perp}\!\otimes\mathbf{\hat{u}}_{\perp}$
with $\otimes$ denoting the dyadic product and the translational-rotational coupling vector
$\mathbf{D}_{\mathrm{C}}=D^{\parallel}_{\mathrm{C}}\mathbf{\hat{u}}_{\parallel}+D^{\perp}_{\mathrm{C}}\mathbf{\hat{u}}_{\perp}$ 
with the center-of-mass position as reference point. The orientation vectors $\mathbf{\hat{u}}_{\parallel}=(\cos\phi,\sin\phi)$ and $\mathbf{\hat{u}}_{\perp}=(-\sin\phi,\cos\phi)$
are fixed in the body frame of the particle. (Notice that eqs.\ \eqref{eq:passiveLangevin} could equivalently be written with friction coefficients instead of diffusion coefficients \cite{MakinoD2004}. In this article we use the given representation since the diffusion coefficients can directly be obtained from our experiments.)
Finally, $\boldsymbol{\zeta}_{\mathbf{r}}(t)$
and $\zeta_{\phi}(t)$ are Gaussian noise terms of zero mean and
variances $\langle\boldsymbol{\zeta}_{\mathbf{r}}(t_{1})\otimes\boldsymbol{\zeta}_{\mathbf{r}}(t_{2})\rangle
=2\:\!\DT \:\!\delta(t_{1}-t_{2})$,
$\langle\boldsymbol{\zeta}_{\mathbf{r}}(t_{1})\:\!\zeta_{\phi}(t_{2})\rangle
=2\:\!\mathbf{D}_{\mathrm{C}}\:\!\delta(t_{1}-t_{2})$, and
$\langle\zeta_{\phi}(t_{1})\:\!\zeta_{\phi}(t_{2})\rangle=2\:\!D_{\mathrm{R}}\:\!\delta(t_{1}-t_{2})$.
Within this formalism, the specific particle shape enters via the translational diffusion coefficients $D_{\parallel}$, $D^{\perp}_{\parallel}$, and $D_{\perp}$, the coupling coefficients $D^{\parallel}_{\mathrm{C}}$ and
$D^{\perp}_{\mathrm{C}}$, and the rotational diffusion coefficient $D_{\mathrm{R}}$.

\subsection{Langevin equations for an active particle under gravity}%
\begin{figure} 
\centering
\includegraphics[width=\linewidth]{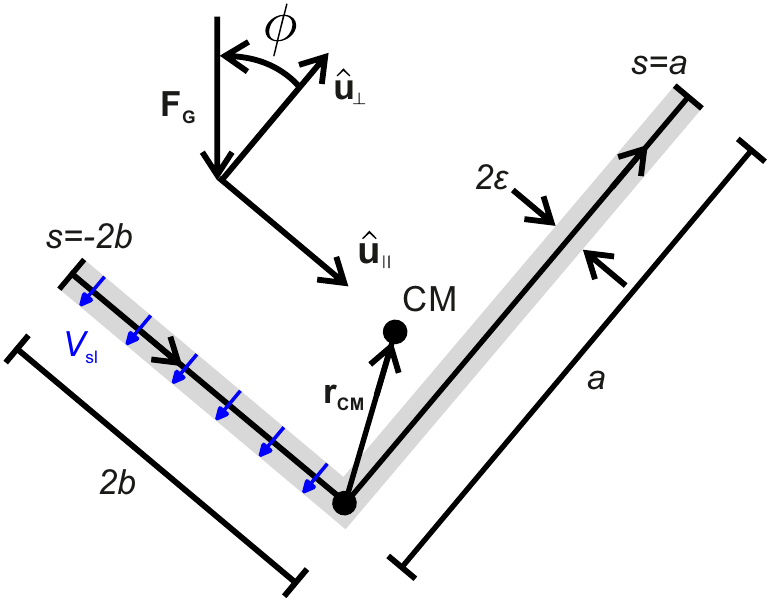}
\caption{\label{Fig.5}\AU{Sketch for the slender-body model.} Illustration of a rigid L-shaped particle with arm lengths $a$ and $2b$. The position $\mathbf{r}(t)$ of the center of mass CM and the orientation angle $\phi(t)$ evolve over time due to the gravitational force $\mathbf{F}_{\mathrm{G}}$ and fluid slip velocity $V_\mathrm{sl}$ on the particle surface. The unit vectors $\mathbf{\hat{u}}_\parallel$ and $\mathbf{\hat{u}}_\perp$ indicate the particle-fixed coordinate system.}
\end{figure} 
In the following, we provide a hydrodynamic derivation of our equations of motion \eqref{eq:Langevinr} and \eqref{eq:Langevinphi} to justify their validity. Our derivation is based on slender-body theory for Stokes flow \cite{Batchelor}. This approach has been applied successfully to model, e.g., flagellar locomotion \cite{Lighthill,Lauga09}. The theory is adapted to describe the movement of a rigid L-shaped particle in three-dimensional Stokes flow as sketched in Fig.\ \ref{Fig.5}. A key assumption is that the width $2\epsilon$ of the arms of the particle is much smaller than the total arc length $L=a+2b$ of the particle, where $a$ and $2b$ are its arm lengths. Though $\epsilon=\unit{1.5}{\micro\metre}$ is only one order of magnitude smaller than $L=\unit{15}{\micro\metre}$ in the experiments, the slenderness approximation $\epsilon\ll L$ offers valuable fundamental insight into the effects of self-propulsion and gravity on the particle's motion, as examined below. 

The centerline position $\mathbf{x}(s)$ of the slender L-shaped particle is described by a parameter $s$ with $-2b\leq s\leq a$:  
\begin{eqnarray}
\label{centerline}
\mathbf{x}(s)= \mathbf{r}-\mathbf{r}_{\mathrm{CM}}+\begin{cases}
s\mathbf{\hat{u}}_\parallel\,, & \text{if\;} -2b\le s\le 0\,, \\
s\mathbf{\hat{u}}_\perp\,, & \text{if\;\,} 0< s\le a\,. 
\end{cases}
\end{eqnarray}
Here, $\mathbf{r}$ is the center-of-mass position of the particle, $\mathbf{\hat{u}}_\parallel$ and $\mathbf{\hat{u}}_\perp$ are unit vectors defining the particle's frame of reference, and  $\mathbf{r}_{\mathrm{CM}}=(a^2\mathbf{\hat{u}}_\perp-4b^2\mathbf{\hat{u}}_\parallel)/(2L)$ is a vector from the point where the two arms meet at right angles to the center of mass CM (see Fig.\ \ref{Fig.5}). The fluid velocity on the particle surface is approximated by $\mathbf{\dot{x}}+\mathbf{v}_\mathrm{sl}(s)$, where $\mathbf{v}_\mathrm{sl}(s)$ is a prescribed slip velocity averaged over the intersection line of the particle surface and the plane through the point $\mathbf{x}(s)$ adjacent to the centerline. Motivated by the slip flow generated near the Au coating in the experiments, we set $\mathbf{v}_\mathrm{sl}=-V_\mathrm{sl}\mathbf{\hat{u}}_\perp$ along the shorter arm ($-2b\le s\le 0$) and no slip (i.e., $\mathbf{v}_\mathrm{sl}=\mathbf{0}$) along the other arm. According to the leading-order terms in slender-body theory \cite{Batchelor}, the fluid velocity is related to the local force per unit length $\mathbf{f}(s)$ on the particle surface and given by
\begin{equation}
\mathbf{\dot{x}}+\mathbf{v}_\mathrm{sl}=c (\mathbf{\underline{I}}+\mathbf{x'}\!\!\otimes\!\mathbf{x'})\mathbf{f}
\label{eq:fluidv}
\end{equation}
with the constant $c=\log(L/\epsilon)/(4\pi\eta$), the viscosity of the surrounding fluid $\eta$, the identity matrix $\mathbf{\underline{I}}$, the vector $\mathbf{x'}=\partial\mathbf{x}/\partial s$ that is locally tangent to the centerline, and the dyadic product $\otimes$. The force density $\mathbf{f}$ satisfies the integral constraints that the net force on the particle is the gravitational force,
\begin{equation}
\mathbf{F}_\mathrm{G}=\int_{-2b}^a \!\!\!\!\!\!\!\mathbf{f}\,\dif s \;,
\label{force}%
\end{equation}
and that the net torque relative to the center of mass of the particle vanishes,
\begin{equation}
\begin{split}%
&\mathbf{\hat{e}}_z\!\cdot\!\!\int_{-2b}^a \!\!\!\!\!\! (-\mathbf{r}_{\mathrm{CM}}+s\mathbf{x'})\!\times\!\mathbf{f}\,\dif s \\
&=\int_{-2b}^0 \!\!\!\!\!\!\! s\mathbf{\hat{u}}_\perp\!\!\cdot\!\mathbf{f}\,\dif s
-\!\int_{0}^a \!\!\!\! s\mathbf{\hat{u}}_\parallel\!\cdot\!\mathbf{f}\,\dif s \\
&\quad\,+\frac{1}{2L}(a^2\mathbf{\hat{u}}_\parallel\!\cdot\!\mathbf{F}_\mathrm{G}+4b^2\mathbf{\hat{u}}_\perp\!\!\cdot\!\mathbf{F}_\mathrm{G})\\
&=0\;,
\end{split}%
\label{torque}%
\end{equation}
where $\mathbf{\hat{e}}_z$ is the unit vector parallel to the $z$ axis. Thus, in 
the absence of gravity the self-propelled particle is force-free and torque-free as required. 

We differentiate eq.\ \eqref{centerline} with respect to time and insert it into eq.\ \eqref{eq:fluidv}. 
This leads to
\begin{equation}
\begin{split}%
&\mathbf{\dot{r}}+\mathbf{v}_\mathrm{sl}+\frac{\dot{\phi}}{2L}(a^2\mathbf{\hat{u}}_\parallel+4b^2\mathbf{\hat{u}}_\perp) \\
&= \begin{cases}%
c(\mathbf{\underline{I}}+\mathbf{\hat{u}}_\parallel\!\otimes\!\mathbf{\hat{u}}_\parallel)\mathbf{f}-s\dot{\phi}\mathbf{\hat{u}}_\perp\,, & \text{if\;} -2b\le s\le 0\,, \\
c(\mathbf{\underline{I}}+\mathbf{\hat{u}}_\perp\!\!\otimes\!\mathbf{\hat{u}}_\perp)\mathbf{f}+s\dot{\phi}\mathbf{\hat{u}}_\parallel\,, & \text{if\;\,} 0< s\le a\,.
\end{cases}%
\end{split}%
\label{slender}%
\end{equation}
The $\mathbf{\hat{u}}_\parallel$ and $\mathbf{\hat{u}}_\perp$ components of eq.\ \eqref{slender} are
\begin{equation}
\begin{split}%
\mathbf{\hat{u}}_\parallel\!\cdot\!(\mathbf{\dot{r}}+\mathbf{v}_\mathrm{sl})
+\frac{a^2\dot{\phi}}{2L} = \begin{cases}
2c\mathbf{\hat{u}}_\parallel\!\cdot\!\mathbf{f}\,, & \text{if\;} -2b\le s\le 0\,, \\
c\mathbf{\hat{u}}_\parallel\!\cdot\!\mathbf{f}+s\dot{\phi}\,, & \text{if\;\,} 0< s\le a\,, 
\end{cases}
\end{split}%
\raisetag{4ex}%
\label{n}%
\end{equation}
and
\begin{equation}
\begin{split}%
\mathbf{\hat{u}}_\perp\!\!\cdot\!(\mathbf{\dot{r}}+\mathbf{v}_\mathrm{sl})
+2\frac{b^2\dot{\phi}}{L} = \begin{cases}
c\mathbf{\hat{u}}_\perp\!\!\cdot\!\mathbf{f}-s\dot{\phi}\,, & \text{if\;} -2b\le s\le 0\,, \\
2c\mathbf{\hat{u}}_\perp\!\!\cdot\!\mathbf{f}\,, & \text{if\;\,} 0< s\le a\,.
\end{cases}
\end{split}%
\raisetag{4ex}%
\label{t}%
\end{equation}
After integrating eqs.\ \eqref{n} and \eqref{t} over $s$ from $-2b$ to $0$ and separately from $0$ to $a$ 
the unknown $\mathbf{f}$ can be eliminated using eqs.\ \eqref{force} and \eqref{torque}. This results in  
the system of ordinary differential equations
\begin{equation}
\begin{split}%
\eta \mathbf{\underline{H}} \begin{pmatrix}
\mathbf{\hat{u}}_\parallel\!\cdot\!\mathbf{\dot{r}} \\
\mathbf{\hat{u}}_\perp\!\!\cdot\!\mathbf{\dot{r}} \\
\dot{\phi}
\end{pmatrix}
\!\!&=\!
\frac{1}{2c}\!\!\begin{pmatrix}
2(a+b) & 0 & -a^2b /L \\
0 & a+4b & -2ab^2 /L \\
-a^2b/L & -2ab^2 /L & A
\end{pmatrix} \!\!\!
\begin{pmatrix}
\mathbf{\hat{u}}_\parallel\!\cdot\!\mathbf{\dot{r}} \\
\mathbf{\hat{u}}_\perp\!\!\cdot\!\mathbf{\dot{r}} \\
\dot{\phi}
\end{pmatrix} \\
&=\begin{pmatrix}
\mathbf{\hat{u}}_\parallel\!\cdot\!\mathbf{F}_\mathrm{G} \\
\mathbf{\hat{u}}_\perp\!\!\cdot\!\mathbf{F}_\mathrm{G} \\
0
\end{pmatrix} + 
\begin{pmatrix}
0 \\
2bV_\mathrm{sl}/c \\
-2 ab^2 V_\mathrm{sl}/(cL)
\end{pmatrix}
\end{split}%
\raisetag{10ex}%
\label{eq:sprop}%
\end{equation}%
with the constant $A=((8L^2-6ab)(a^3+8b^3)-6L(a^4+16b^4))/(12L^2)$. It is important to note that the matrix $\mathbf{\underline{H}}$ is identical to the grand resistance matrix (also called ``hydrodynamic friction tensor'' \cite{Kraft13}) for a passive particle (see further below). 
Since eq.\ \eqref{eq:sprop} is given in the particle's frame of reference, we next transform it to the laboratory frame by means of 
the rotation matrix 
\begin{eqnarray}
\mathbf{\underline{M}}(\phi)=
\begin{pmatrix}
\cos \phi & - \sin \phi & 0 \\
\sin \phi & \cos \phi & 0 \\
0 & 0 & 1
\end{pmatrix}.
\end{eqnarray}
As obtained from the hydrodynamic derivation, the generalized diffusion tensor 
\begin{equation}
\mathbf{\underline{D}}=\frac{1}{\beta\eta}\mathbf{\underline{H}}^{-1} = \left(\begin{array}{ccc}
D_{\parallel} &  D_{\parallel}^{\perp} & D^{\parallel}_{\mathrm{C}} \\
D_{\parallel}^{\perp} & D_{\perp} & D^{\perp}_{\mathrm{C}} \\
D^{\parallel}_{\mathrm{C}} & D^{\perp}_{\mathrm{C}} & D_{\mathrm{R}}
\end{array}\right),
\end{equation}
where $\beta=1/(k_{\mathrm{B}}T)$ is the inverse effective thermal energy, is determined by the parameters
\begin{align}%
D_\parallel &= \left(\!(a+4b)A-\frac{4a^2 b^4}{L^2}\right)\!K\;, \\
D_\parallel^\perp &= \frac{2 a^3b^3 K}{L^2}\;, \\
D_\perp &= \left(2(a+b)A-\frac{a^4 b^2}{L^2}\right)\!K\;, \\
D_\mathrm{C}^\parallel &= \frac{a^2b(a+4b)K}{L}\;, \\
D_\mathrm{C}^\perp &= \frac{4 ab^2 (a+b)K}{L}\;, \\
D_\mathrm{R} &= 2(a+b)(a+4b)K
\end{align}%
with the constant 
\begin{equation}
\begin{split}%
K &= \frac{2c}{\beta}\Big(2(a+b)(a+4b)A \\
&\qquad\;\;\; -\frac{a^2 b^2}{L^2}\big(4abL + a^3 +8b^3\big)\!\Big)^{-1} .
\end{split}%
\end{equation}
Defining 
\begin{align}%
P^* &= \frac{2\beta b^2V_\mathrm{sl}}{c} \;, \\
l &= -\frac{ab}{L} 
\end{align}%
one finally obtains
\begin{align}%
\dot{\mathbf{r}}&= (P^{*}/b) \big(\DT \mathbf{\hat{u}}_{\perp}+l\mathbf{D}_{\mathrm{C}} \big)
+ \beta \DT \mathbf{F}_{\mathrm{G}} \;,\label{eq:Langevinrs}\\
\dot{\phi}&= (P^{*}/b) \big(l D_{\mathrm{R}} +\mathbf{D}_{\mathrm{C}}\!\cdot\!\mathbf{\hat{u}}_{\perp}\big) + \beta \mathbf{D}_{\mathrm{C}}\!\cdot\!\mathbf{F}_{\mathrm{G}}\;,
\label{eq:Langevinphis}%
\end{align}%
where $\DT$ is the translational short-time diffusion tensor, $\mathbf{D}_{\mathrm{C}}$ the translational-rotational coupling vector, and $D_{\mathrm{R}}$ the rotational diffusion coefficient of the particle.
Up to the noise terms, which follow directly from the fluctuation-dissipation theorem and can easily be added, 
eqs.\ \eqref{eq:Langevinrs} and \eqref{eq:Langevinphis} are the same as our equations of motion \eqref{eq:Langevinr} and \eqref{eq:Langevinphi}.
This proves that our equations of motion are physically justified and that they provide an appropriate theoretical framework to understand our experimental results.

Finally, we justify the concept of effective forces and torques by comparing the equations of motion of a self-propelled particle with the equations of motion of a passive particle that is driven by an external force $\mathbf{F}_\mathrm{ext}$, which is constant in the particle's frame of reference, and an external torque $M_\mathrm{ext}$. To derive the equations of motion of an externally driven passive particle, we assume no-slip conditions for the fluid on the entire particle surface. In analogy to the derivation for a self-propelled particle (see further above) one obtains 
\begin{equation}
\begin{split}%
\eta \mathbf{\underline{H}} \begin{pmatrix}
\mathbf{\hat{u}}_\parallel\!\cdot\!\mathbf{\dot{r}} \\
\mathbf{\hat{u}}_\perp\!\!\cdot\!\mathbf{\dot{r}} \\
\dot{\phi}
\end{pmatrix} 
\!\!&= \!
\frac{1}{2c}\!\!\begin{pmatrix}
2(a+b) & 0 & -a^2b /L \\
0 & a+4b & -2ab^2/L \\
-a^2b/L & -2ab^2/L & A
\end{pmatrix} \!\!\!
\begin{pmatrix}
\mathbf{\hat{u}}_\parallel\!\cdot\!\mathbf{\dot{r}} \\
\mathbf{\hat{u}}_\perp\!\!\cdot\!\mathbf{\dot{r}} \\
\dot{\phi}
\end{pmatrix} \\
&=\begin{pmatrix}
\mathbf{\hat{u}}_\parallel\!\cdot\!\mathbf{F}_\mathrm{G} \\
\mathbf{\hat{u}}_\perp\!\!\cdot\!\mathbf{F}_\mathrm{G} \\
0
\end{pmatrix} + 
\begin{pmatrix}
\mathbf{\hat{u}}_\parallel\!\cdot\!\mathbf{F}_\mathrm{ext} \\
\mathbf{\hat{u}}_\perp\!\!\cdot\!\mathbf{F}_\mathrm{ext} \\
M_\mathrm{ext}
\end{pmatrix} .
\end{split}\raisetag{10ex}%
\label{eq:ext}%
\end{equation}
We emphasize that the grand resistance matrix $\mathbf{\underline{H}}$ in eq.\ \eqref{eq:ext} for an externally driven passive particle is identical to that in eq.\ \eqref{eq:sprop} for a self-propelled particle. Formally, the two equations are \textit{exactly} the same if $\mathbf{\hat{u}}_\parallel\!\cdot\!\mathbf{F}_\mathrm{ext} = 0$, $\mathbf{\hat{u}}_\perp\!\!\cdot\!\mathbf{F}_\mathrm{ext} = 2bV_\mathrm{sl}/c$, and $M_\mathrm{ext} = -2 ab^2 V_\mathrm{sl}/(cL)$. This shows that the motion of a self-propelled particle with fluid slip $\mathbf{v}_\mathrm{sl}=-V_\mathrm{sl}\mathbf{\hat{u}}_\perp$ along the shorter arm is identical to the motion of a passive particle driven by a net external force $\mathbf{F}_\mathrm{ext}= F \mathbf{\hat{u}}_\perp = (2b V_\mathrm{sl}/c) \mathbf{\hat{u}}_\perp$ and torque $M_\mathrm{ext} = -( ab/L) F = lF$ with the effective self-propulsion force $F=2b V_\mathrm{sl}/c$ and the effective lever arm $l=-ab/L$. (Note that while the equations of motion of a self-propelled particle and an externally driven passive particle are the same in this case, the flow and pressure fields generated by the particles are different.)

Consequently, the concept of effective forces and torques, where formally external (``effective'') forces and torques that move with the self-propelled particle are used to model its self-propulsion, is justified and can be applied to facilitate the understanding of the dynamics of self-propelled particles.

\subsection{Analysis of the experimental trajectories}
For the interpretation of the experimental results in the context of the theoretical model, it is necessary to determine the  self-propulsion strength $P^{*}$ for the various observed trajectories. This is achieved by means of the relation 
\begin{equation}
P^{*}= b \frac{v_x - \beta F_{\mathrm{G}}(D_{\parallel} - D_{\perp}) \sin\phi \cos\phi}{l D^{\parallel}_{\mathrm{C}} \cos \phi - (D_{\perp} + l D^{\perp}_{\mathrm{C}})\sin \phi} \,,
\label{eq:F}%
\end{equation}
which is obtained from eq.\ \eqref{eq:Langevinr}. It provides $P^{*}$ as a function of the measured center-of-mass velocity $v_{x}$ in $x$ direction. On the other hand, the gravitational force $F_{\mathrm{G}}=mg\sin\alpha$ is directly obtained from the easily adjustable inclination angle $\alpha$ of the experimental setup.

\subsection{Determination of the experimental parameters}
For the comparison of the theory with our measurements the particular values of a number of parameters have to be determined. These are the various diffusion and coupling coefficients, the effective lever arm, and the buoyant mass of the particles.  The translational and rotational diffusion coefficients $D_{\parallel} = \unit{7.2 \times 10^{-3}}{\micro \squaren \metre\reciprocal\second}$, $D_{\perp} = \unit{8.1 \times 10^{-3}}{\micro \squaren \metre\reciprocal\second}$, and $D_{\mathrm{R}} = \unit{6.2 \times 10^{-4}}{\reciprocal\second}$ are obtained experimentally from short-time correlations of the particle trajectories for zero gravity \cite{kummel2013circular}. Since $D_{\parallel}^{\perp}$ is negligible compared to $D_{\parallel}$ and $D_{\perp}$ here, we set $D_{\parallel}^{\perp}=0$. The relation between the two coupling coefficients $D^{\perp}_{\mathrm{C}}$ and $D^{\parallel}_{\mathrm{C}}$ is determined from sedimentation experiments with passive L-particles. The peak position $\phi=\unit{-34}{\degree}$ of the measured probability distribution $p(\phi)$ [see arrow in Fig.\ \ref{Fig.1}(a)] implies $D^{\perp}_{\mathrm{C}} = 0.67 D^{\parallel}_{\mathrm{C}}$ via eq.\ \eqref{eq:phigra}. The only remaining open parameter is the absolute value of the coupling coefficient $D^{\parallel}_{\mathrm{C}}$, which is used as fit parameter in Fig.\ \ref{Fig.2}. Best agreement of the experimental data with the theoretical prediction [see eq.\ \eqref{eq:phipro}] is achieved for $D^{\parallel}_{\mathrm{C}} = \unit{5.7 \times 10^{-4}}{\micro\metre\reciprocal\second}$. The diffusion coefficients have also been calculated numerically by solving the Stokes equation \cite{Carrasco99} and good agreement with the experimental values has been found \cite{kummel2013circular}. The influence of the substrate has been taken into account by applying the Stokeslet close to a no-slip boundary \cite{Blake71}. The effective lever arm of the L-shaped particles is determined to $l=\unit{-0.75}{\micro\metre}$ by assuming an ideally shaped swimmer with homogeneous mass distribution and the effective self-propulsion force acting vertically on the center of the front side of the short arm [see Fig.\ \ref{Fig.1}(d)]. Finally, the buoyant mass $m= \unit{2.5 \times 10^{-14}}{\kilogram}$ is obtained by measuring the sedimentation velocity of passive particles.

\bibliographystyle{naturemag}

\vspace{0.1cm}

\section{Acknowledgements}
This work was supported by the Marie Curie-Initial Training Network Comploids funded by the European Union Seventh Framework Program (FP7), by the ERC Advanced Grant INTERCOCOS (Grant No.\ 267499), and by EPSRC (Grant No.\ EP/J007404).
R.W. gratefully acknowledges financial support through a Postdoctoral Research Fellowship (WI 4170/1-1) from the Deutsche Forschungsgemeinschaft (DFG).

\section{Author contributions}
B.t.H., F.K., R.W., H.L., and C.B. designed the research, analyzed the data, and wrote the paper; 
B.t.H. and R.W. carried out the analytical calculations and the simulations;
F.K. performed the experiments;
D.T. provided the hydrodynamic derivation in the Methods section.

\section{Additional information}
The authors declare no competing financial interests.

\end{document}